# PRAXIS: low thermal emission high efficiency OH suppressed fibre spectrograph

Robert Content*[a], Joss Bland-Hawthorn[b], Simon Ellis[a], Luke Gers[a], Roger Haynes[c], Anthony Horton[a], Jon Lawrence[a], Sergio Leon-Saval[b], Emma Lindley[b], Seong-Sik Min[b], Keith Shortridge[a], Nick Staszak[a], Christopher Trinh[b], Pascal Xavier[a], Ross Zhelem[a]

[a]Australian Astronomical Observatory, PO Box 915, North Ryde, NSW 2113, Australia; [b]Sydney Institute for Astronomy, School of Physics A28, University of Sydney, NSW 2006, Australia; [c]Leibniz-Institut für Astrophysik Potsdam, An der Sternwarte 16, 14482 Potsdam, Germany

## ABSTRACT

PRAXIS is a second generation instrument that follows on from GNOSIS, which was the first instrument using fibre Bragg gratings for OH suppression to be deployed on a telescope. The Bragg gratings reflect the NIR OH lines while being transparent to the light between the lines. This gives in principle a much higher signal-noise ratio at low resolution spectroscopy but also at higher resolutions by removing the scattered wings of the OH lines. The specifications call for high throughput and very low thermal and detector noise so that PRAXIS will remain sky noise limited even with the low sky background levels remaining after OH suppression. The optical and mechanical designs are presented. The optical train starts with fore-optics that image the telescope focal plane on an IFU which has 19 hexagonal microlenses each feeding a multi-mode fibre. Seven of these fibres are attached to a fibre Bragg grating OH suppression system while the others are reference/acquisition fibres. The light from each of the seven OH suppression fibres is then split by a photonic lantern into many single mode fibres where the Bragg gratings are imprinted. Another lantern recombines the light from the single mode fibres into a multi-mode fibre. A trade-off was made in the design of the IFU between field of view and transmission to maximize the signal-noise ratio for observations of faint, compact objects under typical seeing. GNOSIS used the pre-existing IRIS2 spectrograph while PRAXIS will use a new spectrograph specifically designed for the fibre Bragg grating OH suppression and optimised for 1.47 μm to 1.7 μm (it can also be used in the 1.09 μm to 1.26 μm band by changing the grating and refocussing). This results in a significantly higher transmission due to high efficiency coatings, a VPH grating at low incident angle and optimized for our small bandwidth, and low absorption glasses. The detector noise will also be lower thanks to the use of a current generation HAWAII-2RG detector. Throughout the PRAXIS design, from the fore-optics to the detector enclosure, special care was taken at every step along the optical path to reduce thermal emission or stop it leaking into the system. The spectrograph design itself was particularly challenging in this aspect because practical constraints required that the detector and the spectrograph enclosures be physically separate with air at ambient temperature between them. At present, the instrument uses the GNOSIS fibre Bragg grating OH suppression unit. We intend to soon use a new OH suppression unit based on multicore fibre Bragg gratings which will allow an increased field of view per fibre. Theoretical calculations show that the gain in interline sky background signal-noise ratio over GNOSIS may very well be as high as 9 with the GNOSIS OH suppression unit and 17 with the multicore fibre OH suppression unit.

## 1. INTRODUCTION

From 700 nm to 2300 nm, the sky background is mostly made of a large number of narrow OH lines[1] emitted by a layer at an altitude of about 90 km. An optical system that would remove these lines but not the light between the lines would considerably reduce the parasite light in this region.

*robert.content@aao.gov.au; phone +61 2 9372 4846

At low resolution where the OH lines are touching, the removal of the OH lines considerably reduces the parasite light along the spectra so increase the signal-noise ratio everywhere. At high resolution, the OH lines scattering by the spectrograph sends parasite light in the region between the lines so removing the OH lines before they enter the spectrograph also removes this parasite light. Also, lines of interest mixed with the OH lines can be made visible if the resolution at which the OH lines are removed is higher than the resolution of the spectrograph and the distance between a line of interest and the adjacent OH line is larger than a little more than half the OH suppression width. For more than 20 years now, optical systems have been developed to remove the OH lines of the sky background in the Near-Infrared (NIR). The first systems were all using a large spectrograph with a mask and a mirror in the spectral plane to remove the OH lines and send the light back to a second grating in the spectrograph that reconstructs the original white light slit image without the OH lines or directly gives a spectrum on a detector. Systems with the reconstructed slit were built for the Subaru telescope[2]. These instruments were highly successful at removing the OH lines while maintaining a high transmission between the lines. The OH suppression optics were followed by a low resolution spectrograph. Spectrographs as CIRPASS and FMOS[3] directly give low or high resolution spectra on a detector. The OH line light scattered between the lines by the spectrograph at high resolution is however not removed. Work on an OH suppressed camera was also done for high signal-noise ratio imaging[4] instead of spectroscopy.

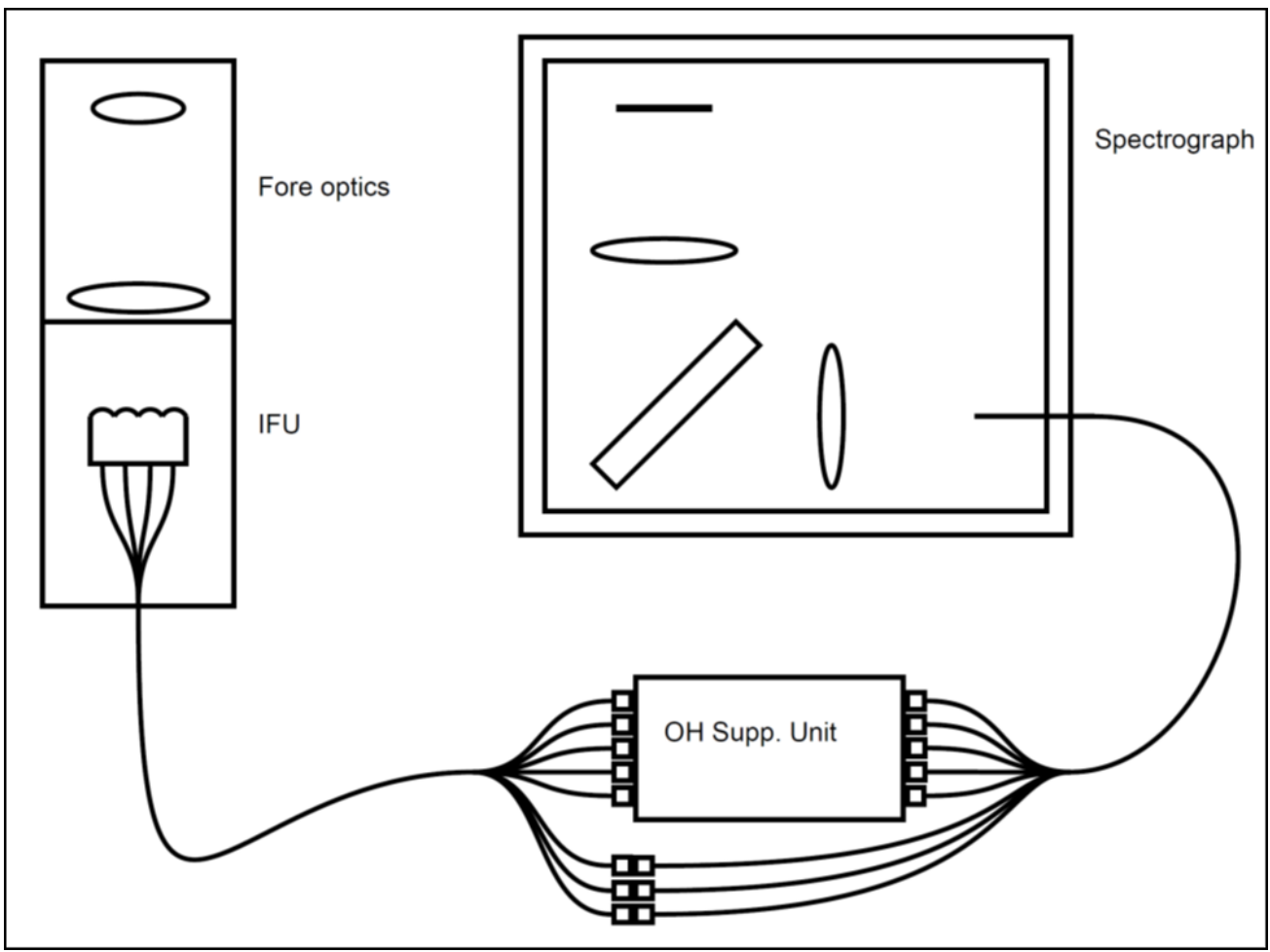

Figure 1. Schematic showing the components of PRAXIS.

Our collaboration has developed another method which uses Bragg gratings in single mode fibres to reflect the OH lines while maintaining a high transmission between the lines[5]. A fibre Bragg grating (FBG) is a length of single mode fibre with refractive index variations along the axis which cause specific wavelengths to be reflected. Aperiodic FBGs can be used to create multi-notch filters with large numbers of narrow, deep, square profiled notches while retaining high interline throughput, exactly the properties required for effective filtering of the OH emission lines that dominate the near infrared sky background. The current state of the art allows for over 100 notches at a resolution of R ~ 10000, depths of over 30dB, and interline throughput of over 90%. The single mode fibres (SMFs) required for FBGs generally cannot be used directly in astronomical applications because of low coupling efficiencies except with extreme adaptive optics where the input PSF diffraction cores are well defined. However, devices known as photonic lanterns (PLs) enable arrays of FBGs (or other single mode photonic devices) to be used with multimode fibres inputs and outputs.

FBGs have the potential to considerably increase the signal-noise ratio in fibre MOS spectrographs and fibre IFUs in the range from 700 nm to 2300 nm. To implement this however, it will be necessary that the spectrographs have very low instrumental noise so that they remain sky background noise limited. This is especially important since the OH suppression system does come with some losses of transmission. Simulations as well as real world experience gained during the commissioning of the GNOSIS system[6] with the IRIS2 spectrograph on the AAT suggest that existing spectrographs, built for significantly higher sky background levels, will be unable to fully exploit the benefits of FBG OH suppression. Because of this and other limitations, a new spectrograph specifically designed to work with FBG systems is needed which is why PRAXIS has been designed and is being built. The FBG is designed for the range of wavelength from 1470 nm to 1700 nm. A future independent system will cover the J band.

## 2. OPTICAL DESIGN

Borrowing the terminology of controlled thermonuclear fusion, GNOSIS "broke even", that is the losses of transmission were approximately compensated by the reduction in background leaving a signal-noise ratio roughly the same than without the OH suppression system. The goal of the new design of spectrograph, accessory optics and mechanical system is to prove the technology by getting a signal-noise ratio larger than would be achieved with a similar system without OH suppression. Building on the experience acquired with GNOSIS, a complete redesign of the system was done to solve the different problems encountered. Special care was especially taken to stop the thermal background from leaking toward the detector. This was made even more difficult due to the need to separate the camera from the spectrograph to reduce cost. Care was also taken in maximizing the transmission and in the choice of a high efficiency low noise detector.

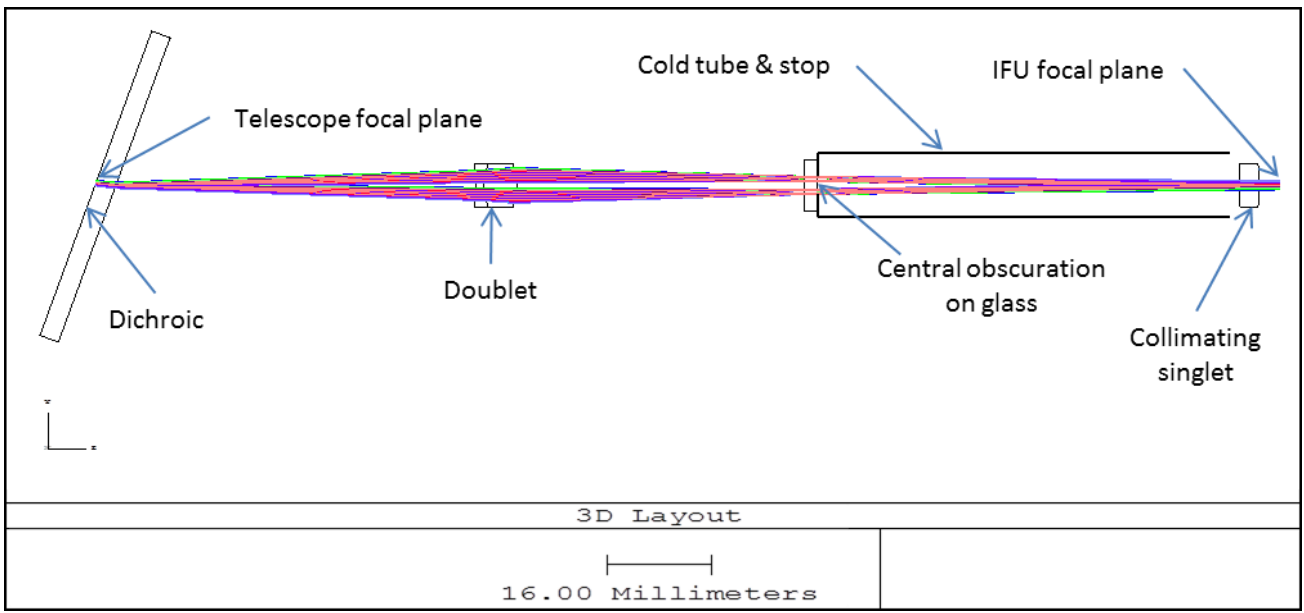

Figure 2. Layout of the fore-optics.

### 2.1 Basic design

The optical train (fig. 1) after the telescope focal plane starts with a dichroic that reflects the visible light to a guiding camera while being transparent to the wavelength range from 1050 nm to 1700 nm. It is followed by the fore-optics that re-images the focal plane on an Integral Field Unit (IFU) with the proper magnification. This IFU is made of a microlens array followed by one fibre per lenslet. Some of these fibres are for reference and acquisition while the others are each followed by a photonic lantern that efficiently distributes the light to an array of single mode fibres. Each of these contains 2 Bragg gratings in series designed to each reflect a different set of OH lines. At the output, another photonic lantern links the single mode fibres to a multi-mode fibre with a core equal in size to the core of the first multi-mode fibre. All the reference and OH suppressed fibres end at the slit of the spectrograph. All optics will be covered by a highly efficient anti-reflection coating (<0.5% reflectivity, goal <0.25%) made possible by the small bandwidth.

## 2.2 Fore-optics

The fore-optics (fig. 2) reimages the focal plane of the telescope on an IFU with the proper magnification. They must also block as much thermal emission as possible from entering the IFU. The optical train is made of a doublet and a singlet with the latter there mostly to collimate the beam before the IFU. This is sufficient to give a high image quality. The main difficulty in the design was to block the thermal emission. The optics were put in a Dewar and a cold stop was added. The doublet is used as the input window and the singlet as the output window so they are at room temperature. They are made of glasses that have a sufficiently low absorption at 1700 nm to avoid thermal emission. A central obscuration printed on a glass plate was also added at the center of the cold stop. At the output, the IFU must only be receiving thermal emission from the cold part of the fore-optics. To do this, a cold tube was added that covers the region from the stop to the singlet and the singlet was oversized. Precise calculations of the needed singlet size were made by modeling the transmission of the IFU and following optics for each direction and position of an incident ray on the IFU and surrounding glass. Any ray that can enter a fibre had to be accounted for. Two interchangeable set of fore-optics will be available for spatial resolutions of 0.55" and 0.8".

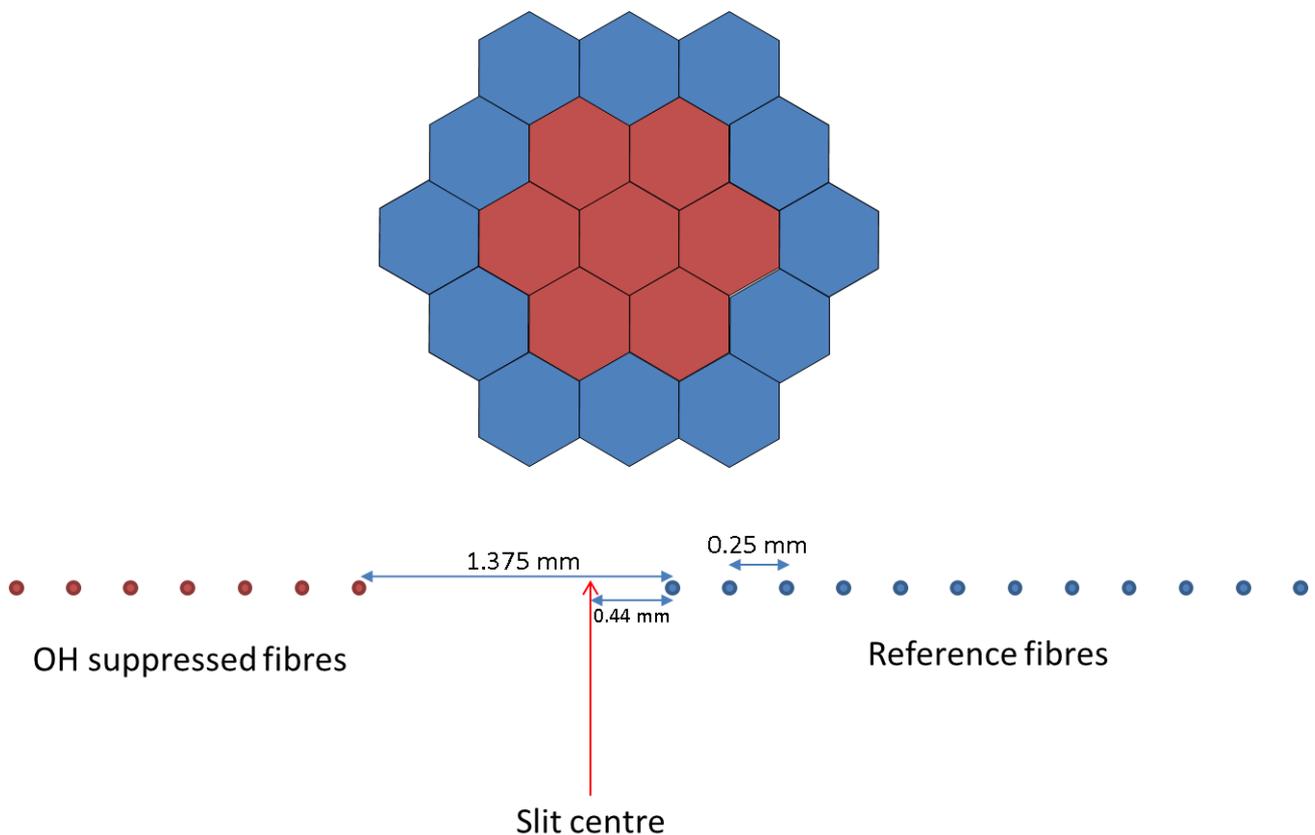

Figure 3. Schematic of the IFU and spectrograph slit. Brown lenslets and fibre ends are attached to an OH suppressed system while the blue ones are not.

## 2.3 Integral field unit

The IFU is made of a microlens array glued to an array of fibre ends (fig. 3). It has 19 hexagonal lenslets of which the 7 in the center are attached to an OH suppression system while the 12 others on the periphery are reference lenslets for measurement of the sky spectrum. They are also used for acquisition and alignment of the object. Two IFUs will be available, one with 0.55" lenslet width, the other 0.8". Detailed modeling of the whole optical train was made to calculate the resulting SNR and determine the optimum microlens size. Each size is optimized for a specific FBG system. The 0.55" will be used with the already available FBG system from GNOSIS. The size of the GNOSIS lenslets were 0.4" but the goal of GNOSIS was to demonstrate the ability to remove the OH lines by observing the sky while maximizing the transmission. The PRAXIS size is optimized to maximize the light captured by the IFU for a small object observed in typical seeing. A trade-off was made between the field of view and the transmission to maximize the total intensity. The 0.8" will be used with a new FBG system in development using multi-core fibres in which each core is equivalent to a single mode fibre.

## 2.4 OH suppressed system

Each of the seven 50 μm fibre fed by a lenslet in the center of the IFU is attached to a photonic lantern that efficiently distributes the light between 19 single mode fibres (0.55" lenslets) or 55 single mode cores (0.8" lenslets). Each single mode fibre or core has 2 FBGs in series. Another photonic lantern at the output recombines the light into another 50 μm fibre. Because of the symmetry between input and output, any light lost in the system will be replaced by a corresponding thermal emission. For example if a proportion of the light is absorbed by the cladding at the input then an emissivity of the same proportion will generate thermal emission from the cladding at the output. Similarly, if part of the light at the input is reflected toward the buffer and absorbed, a symmetrical thermal emission from the buffer will be reflected into the output fibre. To reduce the thermal emissions, we will cool the OH suppression system. The GNOSIS system can be cooled down to -10C. We will put it in a commercial freezer which can be cooled down to -18C. The multi-core system will be designed to be cooled to an even lower temperature.

Each output fibre has its fibre end on the spectrograph slit. To avoid thermal emission from the cladding we planned to cool the last few centimeters of the fibre. However, thermal emission can travel from the hot part of the fibre through the cold part and into the spectrograph if the angle of the rays with respect to the axis of the fibre is sufficiently small. This is due to the difference in refractive index between the cladding and the buffer. The larger the incident angle at the interface cladding-buffer the higher is the reflectivity. Also, the distance between reflections increases which reduces the number of reflections along the cold length. These 2 effects combine to make a hot black body region at the center of the spectrograph stop. The length that needs to be cooled is of the order of 1-m. Note that geometrical optics gives a pessimistic value of this thermal emission. Two wave optics phenomena will reduce this effect. First, the evanescent waves in the buffer will be submitted to absorption. Also, the lowest mode in the center will have a non-zero angular width more or less equivalent in geometrical optics to a minimum angle between rays and the fibre axis.

## 2.5 Spectrograph

The spectrograph is made of transmissive collimator and camera separated by a Volume Phase Holographic (VPH) grating (fig. 4). The detector is separated from the rest of the optics to reduce cost. It permits to use a smaller Dewar and simplify the mechanism for the detector alignments in focus, tilt, position and rotation. To reduce the number of optics, the widows on each side of the separation between spectrograph and detector have power. The spectrograph is designed primarily for the range 1470 to 1700 nm but can be used in the range 1090 to 1260 nm after a change of grating and a refocus. The resolution is 2500 with 2 pixels per FWHM spectral elements. All surfaces are spherical. Glasses were chosen to have a low absorption in the range of interest and to be of reasonably low cost. One exception was the spectrograph window which needed to have a high refractive index but a low absorption coefficient. ZnSe had to be used.

As for the rest of the optical train, thermal emission was the main concern. The spectrograph can have even more leaks of thermal emissions toward the detector than GNOSIS because of the separation between the spectrograph optics and the detector. The spectrograph optics and the detector are both in a Dewar. The spectrograph optics are inside a cold enclosure at -80C. The enclosure of the detector offered by the manufacturer did not sufficiently block the thermal emission. A baffle system and a mask in front of the detector was designed by us and added by the manufacturer. A cold filter behind the baffles cut the wavelengths larger than 1700 nm. The two windows have been oversized to make sure that any thermal emission hitting the detector comes from the cold enclosure at -80C inside the spectrograph. Calculations were performed with ZEMAX to determine the exact size of the windows. One problem of the original design was that the window of the spectrograph was made of a Shott glass with relatively high absorption coefficient which would have resulted in thermal emission since the windows are at ambient temperature. The glass was also of quite high refractive index so it could not be replaced by fused silica. This is why it had to be replaced by ZnSe.

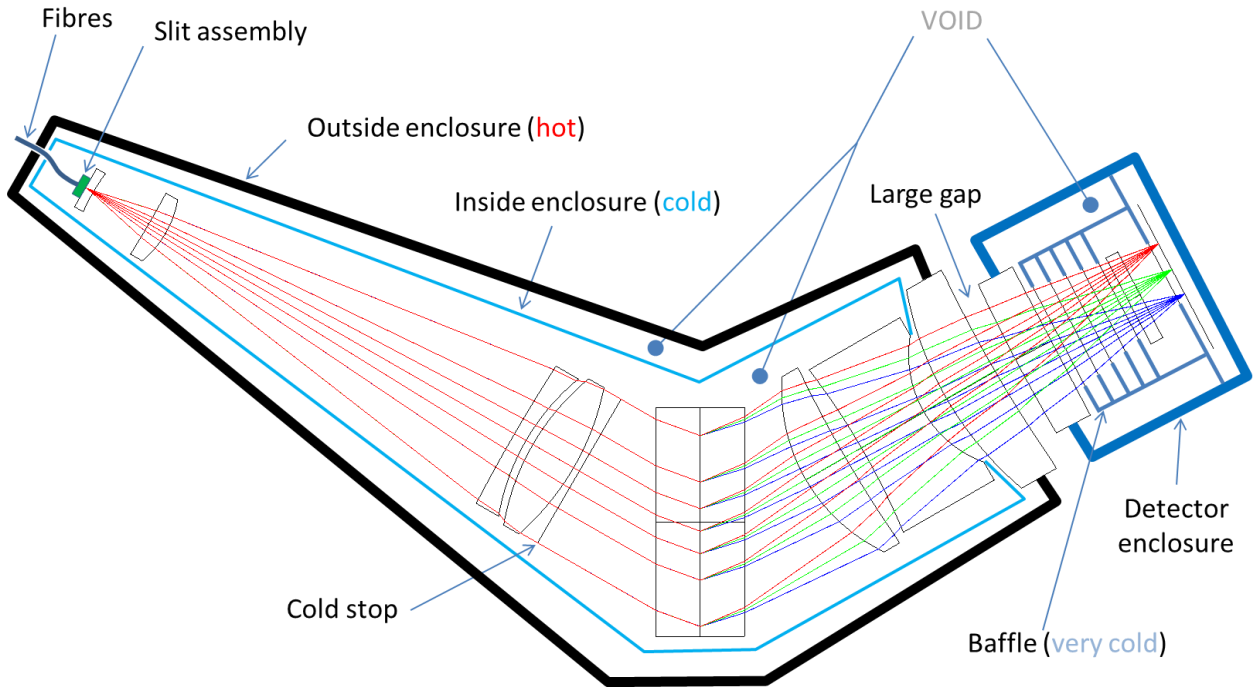

Figure 4. Schematic of the spectrograph and detector enclosures with the ZEMAX layout of the optics.

## 3. OH SUPPRESSION WITH MULTI-CORE FIBRES

The inclusion of multi-core fibre Bragg gratings (MCFBGs) in the OH suppression unit for PRAXIS has significant benefits over the SMF-based grating unit which was used in GNOSIS. It reduces the size, weight and physical complexity of the unit by performing the function of sets of SMFBGs within a single fibre. Using MCFs also reduces the time required to create gratings, as a single UV laser exposure with a phase mask can inscribe all cores at once compared to performing the same task for each fibre separately when using SMFs. No splices are required to incorporate the MCFBGs into the rest of the optics; instead, the ends of the inscribed MCF are tapered into MMFs. This requires less time-consuming manual labour and results in improved throughput. Figure 5 is a diagram of the functional layout of both types of OH suppression units, demonstrating in particular the improved compactness and streamlining provided by MCFBGs.

Work on producing uniform MCFBGs with spectral properties equal to that of SMF gratings is ongoing. A team at the University of Sydney is developing MCFBG writing procedures which allow the inscription of gratings using the same equipment as for SMFBGs[7]. Figure 6 represents our current ability to write MCFBGs with large ($n > 19$) numbers of cores, showing the amount of suppression (in dB) at each core's Bragg wavelength when a 55-core fibre was inscribed with a single notch. The outcomes are better for 7 and 19-core MCFs owing to the greater distance between cores and reduction in self-shadowing, but the processes are still in development.

The writing process must be modified from the standard SMF procedure to ensure all cores within the MCF receive the same exposure, both in terms of the power delivered to each core and the resulting spectral response. Optimum functioning of the MCFBGs relies on every core having the same response to incoming light; even a single core's Bragg wavelength being detuned from the others drastically reduces the suppressing power of the system. Once these issues are resolved, PRAXIS will be able to take advantage of the benefits of MCFBGs in its OH suppression unit.

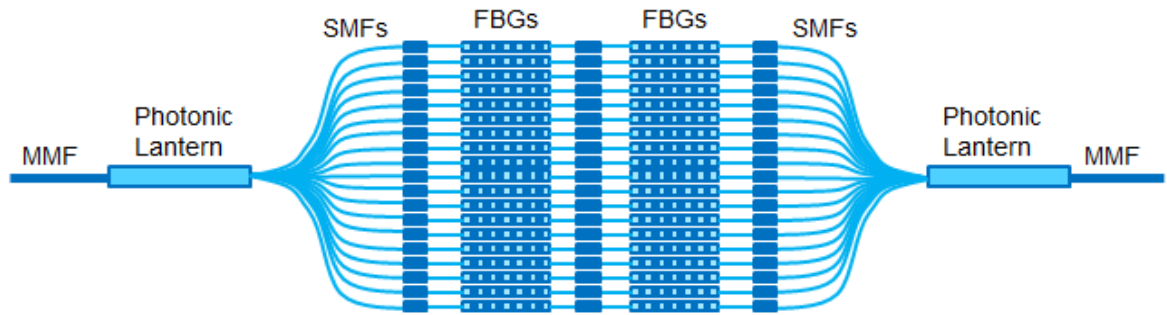

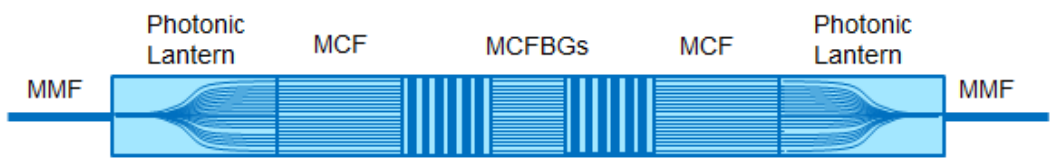

Figure 5. A schematic of the differences between GNOSIS' SMF-based OH suppression unit and its planned MCFBG-based replacement in PRAXIS. Each individual FBG in GNOSIS had to be written and spliced manually; with a total of 133 'tracks' (made up of seven 1 x 19 lanterns) each requiring multiple splices this was time-consuming and resulted in a bulky device once packaged. The MCF version requires only a single exposure for all cores and no splices.

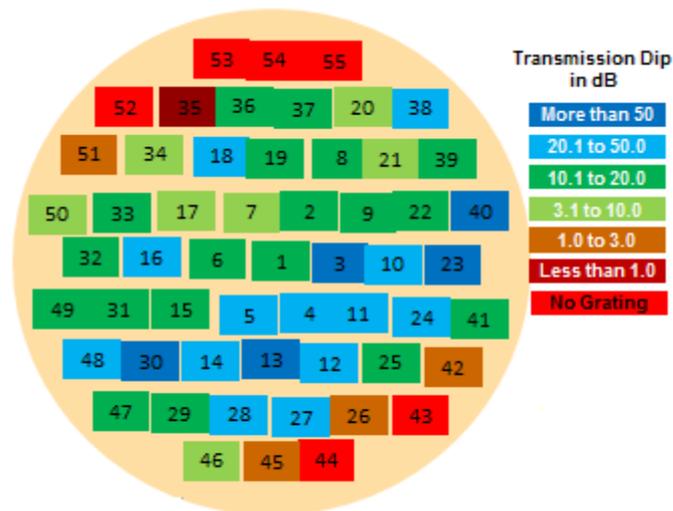

Figure 6. A diagram of the differences in maximum grating suppression for a single notch written into a 55-core MCF with total cladding diameter 230 μm and core diameter 6.5 μm. A polished capillary tube was used to flatten the incoming field and improve on the standard SMF writing process as outlined in [7]. There are also variations in the Bragg wavelength and grating width across the fibre, not pictured here.

# 4. OPTO-MECHANICAL DESIGN

## 4.1 Fore-optics units

The PRAXIS Instrument consists of two interchangeable fore-optics assemblies (fig. 7) that interface to the Anglo-Australian Telescope (AAT) Cassegrain Unit Refurbished Environment (CURE) instrument facility. The CURE facility is an add-on to the AAT's main Cassegrain focal station. It provides an alternative mechanical interface, additional acquisition & guiding (A&G) facilities and additional calibration sources. This facility eliminates the need for AAT instruments to provide their own A&G and calibration facilities and also simplifies instrument exchange. The CURE facility is designed to meet the needs of compact instrumentation with moderate fields of view.

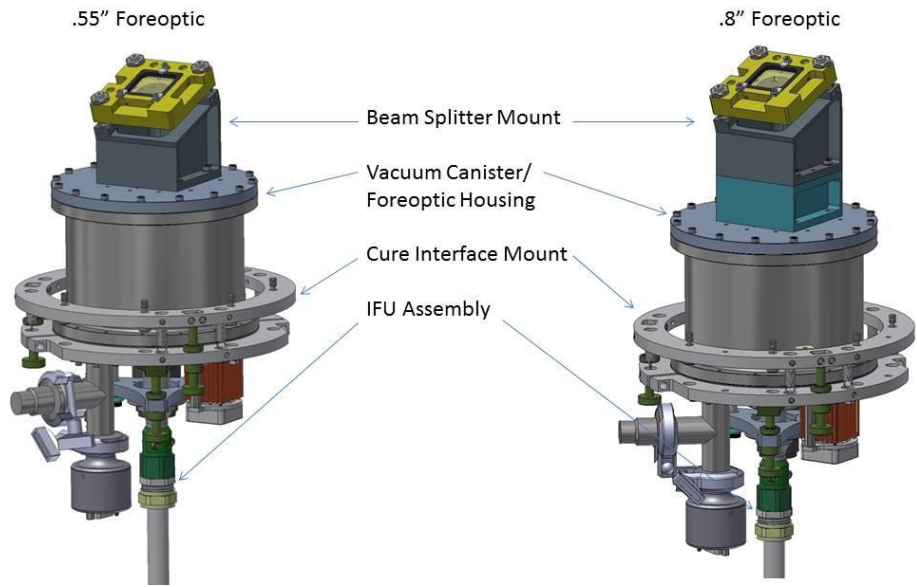

Figure 7. Top Level PRAXIS .55" and .8" Fore-optic Assemblies.

The PRAXIS fore-optics require a cooled cold stop, and cold tube at -20°C to block emission from around the AAT primary and secondary mirrors and any extraneous stray light from entering the IFU. To achieve this requirement, the fore-optics were designed as a Dewar with each lens acting as a window to form a vacuum vessel (fig. 8). The interior of the housing will be evacuated. All components that comprise the vacuum vessel will be constructed from stainless steel. Each lens will be aligned in tip and tilt and bonded within its lens cell with RTV566. The bottom lens will be fixed with a precision bore in the housing. The top lens has provision for alignment in (X,Y) relative to the bottom lens. Mounted internally to the housing is a thermoelectric cooler that will provide the required cooling to the cold stop and cold tube assembly. The thermoelectric cooler is mounted to a copper block that allows waste heat to transfer to the exterior of the housing. Waste heat is removed from the TEC hot side via a forged PIN fin heat sink and impairment airflow via a 35mm fan. The TEC will provide approximately 2W of cooling lift while operating within a maximum dome temperature of 20°C. Steady state analysis shows that only 1.2W of cooling should be required leaving adequate headroom on the cooler. Transient analysis shows that cool down time for the stop and the cold tube assembly should take no longer than 20 minutes to achieve target temperature. All components in the cold path will be constructed of copper and then blackened with Ebanol C black oxide process. The cold path is thermally isolated from the ambient Dewar housing using a G10 spider and stainless steel stand offs providing a high thermal resistivity to ambient conductive losses. The thermoelectric cooler will be sourced from TE Tech and will include their controller, TE-48-20. Pressure level can be monitored via a MKS Instrument925 MicroPirani transducer. Electrical feed through into the vacuum chamber will be sourced from Kurt Lesker. As this is not a facility grade instrument all control and monitoring will be local to the instrument. Vacuum hold time is not of critical importance for the fore-optic and may be pumped down each night. The fore-optic feeds a 19 fibre IFU assembly sourced from Fiber Guide. The IFU fibres are

#AFS50/125/250Y, arranged on a hexagonal pattern with a 250 μm pitch. The cable assembly is protected in a custom designed conduit assembly.

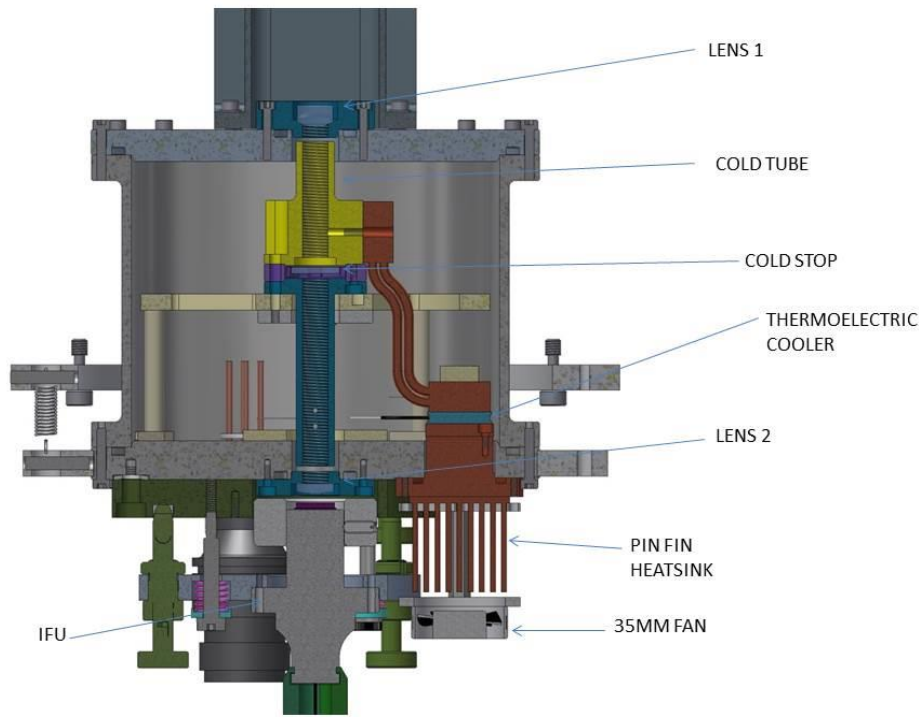

Figure 8. PRAXIS Fore-optics cross section view.

**4.2 Praxis spectrograph assembly**

PRAXIS consists of an AAO designed spectrograph feeding a GL Scientific with Hawaii 2RG detector (2048 x 2048 pixels with 18μm pixel pitch). The spectrograph contains a 19 fiber slit, 4 lens element collimator, swappable H-Band and J-Band Gratings, and 3 lens element camera. The GLS Cryostat is modified with a custom, slightly powered window, custom baffling, and cold filter (fig. 9, 10).

The spectrograph design requires cooled optics with the exception of the exit window at a level of 200K minimum. Cooling is achieved through a Sunpower Cryotel MT Cryocooler. Thermal straps route cooling to each of the optic assemblies, consisting of the grating, collimator, and camera. Overall temperature control is able to be tuned with heaters located at each of the lens assemblies. For stray light elimination all mechanical components are blackened via Ebanol C process and a baffling arrangement is also included. The baffle is cooled, blackened internally and polished externally to act as a radiation shield. Alignment tolerances do not allow a drop fit of the lens assembly so each lens is adjustable in tip-tilt. The lenses are tangentially mounted with a three button arrangement. Careful selection of the housing material and button allow for an athermalized design. The collimator and camera lens housings are mounted with a G10 spider arrangement to isolate the housings from the ambient exterior temperature.

The exterior housing is constructed of 304 Stainless Steel. It consists of 5 major parts. The collimator housing holds the fiber feed through, slit, and collimator. The housing also establishes the optical axis. The camera assembly holds the spectrograph camera doublet, and powered window. The main housing, a 304 forging, joins all the major components, and allows for a swappable grating and stop. The top and bottom plates allow the electrical services to enter through a feedthrough, vacuum valve, and cryocooler mounting. The GLS Cryostat mounts directly to the spectrograph through a custom designed focus, tip and tilt mount, manually adjustable to save cost.

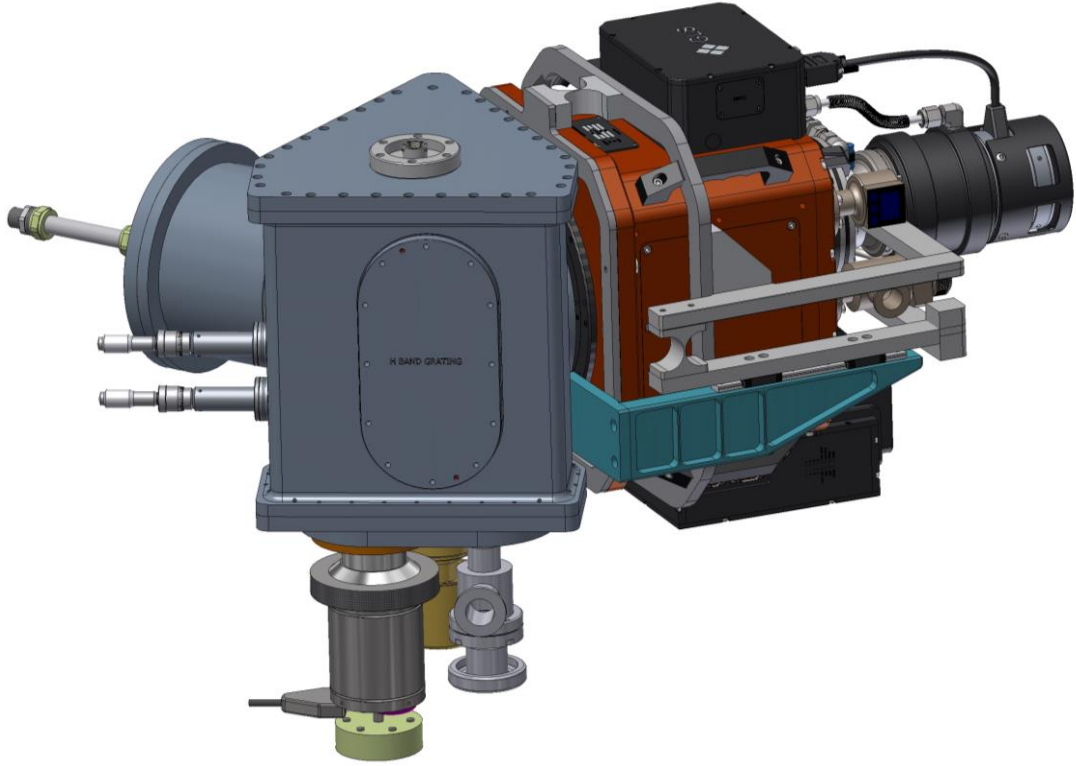

Figure 9. PRAXIS Spectrograph top level assembly.

## 5. SIMULATIONS

### 5.1 Thermal background

Observations with GNOSIS[6,8] showed that although strong OH suppression was achieved by the FBGs, the overall sensitivity of IRIS2 was not improved. This was due to two reasons: the low overall throughput of GNOSIS + IRIS2 meant that below 1.6 µm the background was dominated by the detector dark current, and above this the background was dominated by thermal emission from the warm relay optics.

PRAXIS is designed to ensure that the emission from the instrument does not exceed that of the telescope itself. To this end the fore-optics are baffled and cooled, including a cold-stop with a central obstruction; the entire grating unit including the photonic lanterns will be cooled; the fibres entire the spectrograph Dewar via a vacuum feed-through; the spectrograph Dewar is cooled; the detector Dewar is cryogenically cooled and baffled.

To determine the required temperature for each component we have modelled the thermal emission for the entire system. Each component of the instrument was modelled as a greybody, with a given temperature, $T$, and emissivity, $\varepsilon$. Thus the background, $N$, in ph s$^{-1}$ m$^{-2}$ µm$^{-1}$ arcsec$^{-2}$ is given by (chapter 7, Allen's Astrophysical Quantities),

$$N = \frac{1.41 \times 10^{16} \varepsilon}{\lambda^4 \left( e^{\frac{14387.7}{\lambda T}} - 1 \right)},$$

where $\lambda$ is in µm and $T$ is in K.

The black-body spectrum for each component is then multiplied by the combined efficiency of all components downstream, and the minimum $A\Omega$ product of the downstream components, to give a spectrum in ph s$^{-1}$ µm$^{-1}$. This is then binned into pixels on the detector, taking into account the modelled dispersion, to yield a spectrum in ph s$^{-1}$ pix$^{-1}$ for each component. The required temperature for each component can then be calculated by adjusting the appropriate value of *T* within the model, such that the contribution is not significant, or minimized as much as possible. The emissivity of each component is generally taken to be one minus the throughput, except in the case of the photonic lanterns, which have an angular throughput dependence, which must be calculated in the far-field, and integrated over the NA of the multi-mode fibres. Additionally, there is an extra component of loss at the front face of the photonic lanterns, since the MMFs can accept thermal emission over their entire NA, but only 19 modes are coupled to the lanterns; this only affects the thermal emission from the microlens array itself; the emission from upstream components is fed to the fibres at a slower speed, such that only ~ 19 modes are excited.

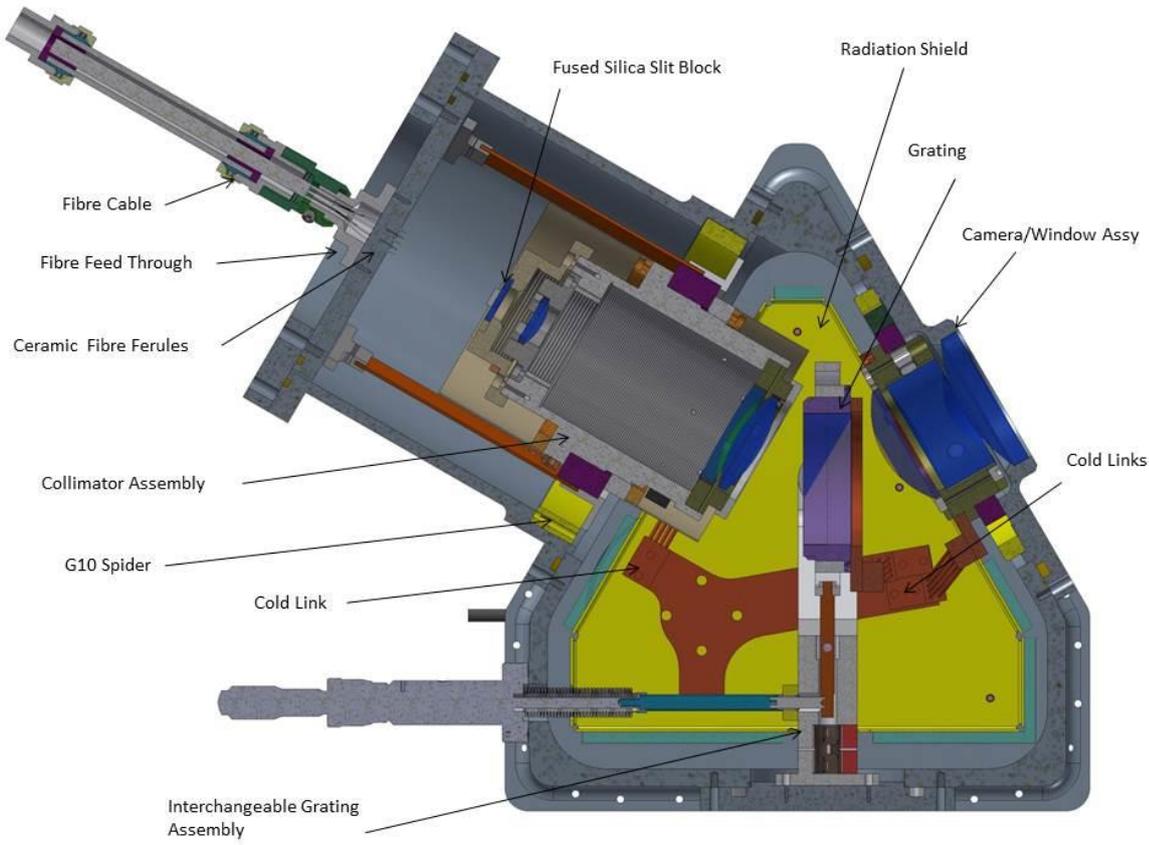

Figure 10. PRAXIS Spectrograph Cross Section.

Table 1. Required temperatures of the various components of the instrument.

| Component | Temperature (K) |
|---|---|
| Fore-optics | Dome temp. (273 – 293) |
| Fore-optics baffle and cold-stop | 253 |
| IFU | Dome temp. (273 – 293) |
| Grating unit | 263 |
| Spectrograph Dewar | 203 |
| Detector Dewar and filter | 130 |

Exceptions to this procedure occur for any component that is downstream of the grating, and for any light redder than the cut-off of the filter, emission from which will be seen by the entire detector.

The final required temperature for each component is given in Table 1, and the spectra of each component, as well as the total thermal spectrum are shown in Figure 11.

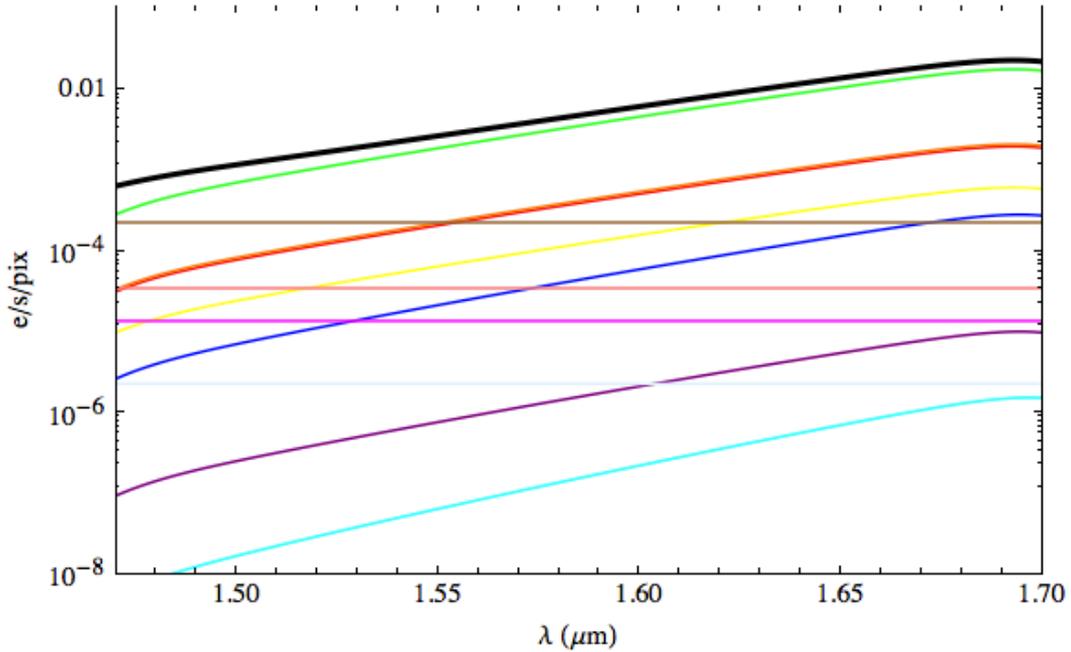

Figure 11. The total expected thermal background of GNOSIS (thick black line), and the individual spectra of the various components: primary mirror (red), secondary mirror (orange, superimposed to red), cold-stop (blue), IFU (yellow), fore-optics baffle scattered light (purple), grating unit (green), slit (cyan), zeroth order image (magenta), spectrograph window (brown), grating reflections (light blue), filter (pink).

### 5.2 Photonic lanterns

Photonic lanterns are an important enabling technology for astrophotonics with a wide range of potential applications including fibre Bragg grating OH suppression, integrated photonic spectrographs and fibre scramblers for high resolution spectroscopy. The behavior of photonic lanterns differs in several important respects from the conventional fibre systems more frequently used in astronomical instruments and a detailed understanding of this behavior is required in order to make the most effective use of this promising technology. To this end we have undertaken a laboratory study of photonic lanterns with the aim of developing an empirical model for the mapping from input to output illumination distributions. We have measured overall transmission and near field output light distributions as a function of input angle of incidence for photonic lanterns with between 19 and 61 cores. A full description of this work can be found in Horton et al.

The most important measurements needed to simulate the whole system were the transmission of the photonic lantern as a function of incident angle. With a multimode fibre, the transmission is not far from being a top hat while it is nearly Gaussian with a single mode fibre. Measurements were done on a photonic lantern of the OH suppression system of GNOSIS which we will use at first with PRAXIS (fig. 12). It transfers light from a 50 μm core of a multimode fibre to 19 single mode fibres. The result shows a transmission intermediate between a multimode and a single mode fibre. The center is flatter than a Gaussian but far from the near top hat shape of a multimode fibre. Also, the wings are shorter than a Gaussian. Similar results were obtained with the photonic lanterns made for transferring light from a 100 μm core of a multimode fibre to 61 single mode fibres. These measurements were used to extrapolate the transmission from a 50 μm core to a 55 cores multicore fibre which we intend to use in the second phase of the PRAXIS project.

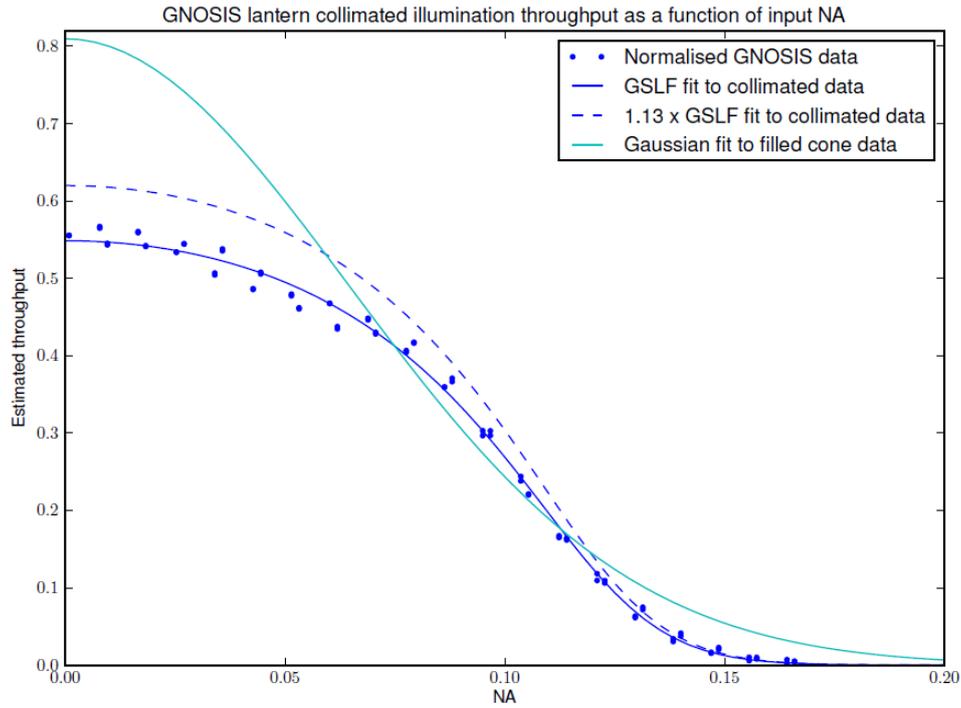

Figure 12. Throughput of the GNOSIS photonic lantern with Thorlabs M16L01 SMA/FC fibres (50 μm core) relative to the M16L01 fibres alone. The throughput is for collimated illumination and is plotted as a function of the numerical aperture corresponding to the angle of incidence of the input illumination.

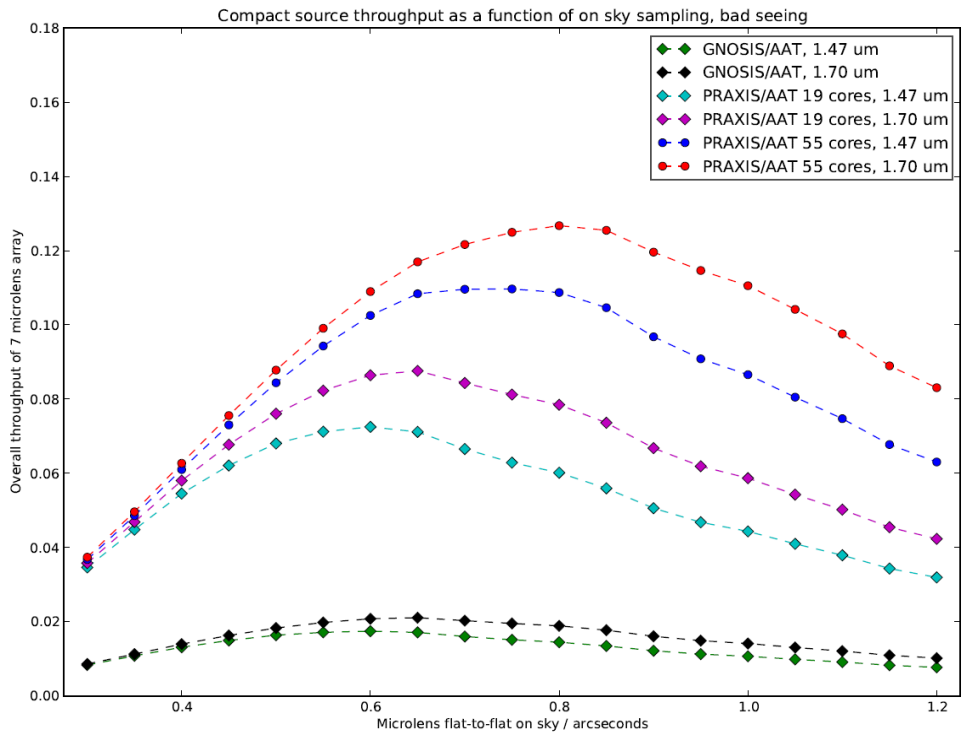

Figure 13. Overall system throughput for compact source centred on an array of 7 photonic lanterns fed by hexagonal microlenses versus the on sky flat-to-flat field of view per microlens. The atmospheric seeing is assumed to be 1.67" at λ = 500nm which equates to 1.33" at λ = 1.585 μm.

## 5.3 Whole of PRAXIS

The results from the tests of the photonic lantern and the thermal emissions were used to calculate the end to end throughput and the signal-noise ratio. See Horton et al. for a detailed analysis. It was used in particular to calculate the optimum size for the microlenses. If the size of the microlenses is increased, they capture more light over the field of view of 7 microlenses but their transmission is reduced. The throughput is the product of the two so a trade-off must be made. Also, we are interested in observing point source objects like stars but also extended objects like the sky background itself. The optimum size is not the same for these two so another trade-off is necessary. For a point source, the throughput increases with microlens size at first because more light is capture while the transmission is reduced slowly but with large size there is little more light captured by the field while the transmission is rapidly reduced. The throughput first increases, passes by a maximum then decreases (fig. 13). The situation is different with an extended source. The throughput continually increases with microlens size and tends asymptotically to a maximum at infinity (fig. 14). As already mentioned, we settled for a microlens width of 0.55" for the GNOSIS FBG and 0.8" for the multicore FBG. The increase in signal-noise ratio over the GNOSIS microlens width of 0.4" seems as high as 9 with PRAXIS using the same OH suppression system and 17 using the new multicore fibres.

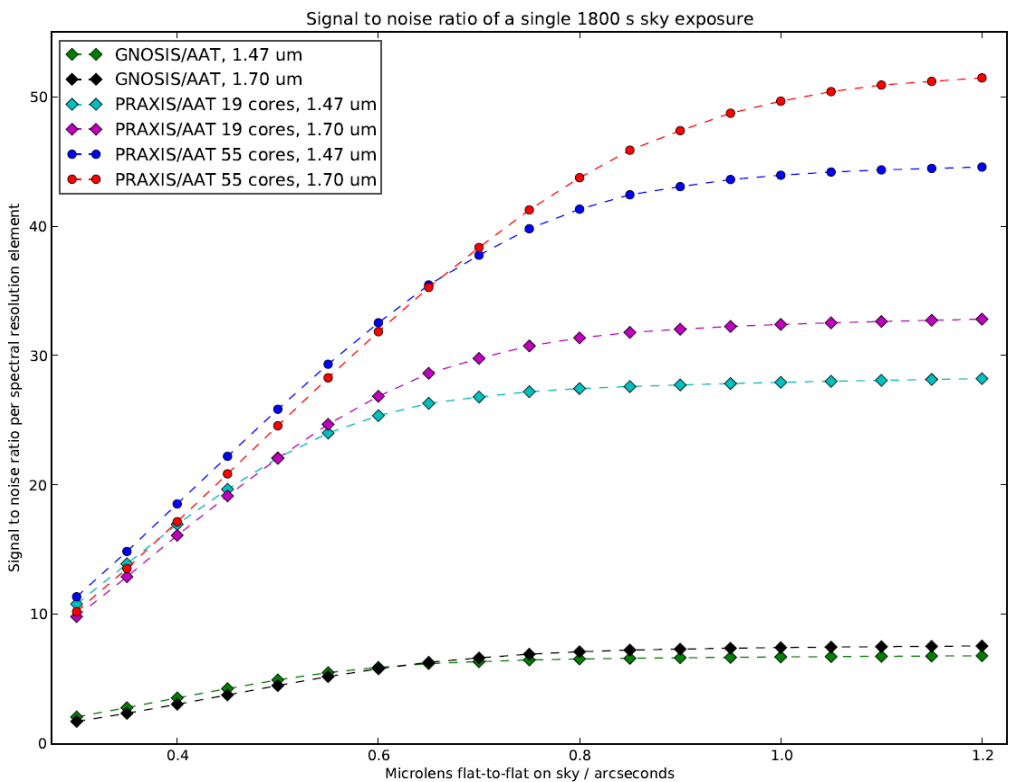

Figure 14. Predicted signal to noise ratio per spectral resolution element for the summed sky spectra of an array of 7 photonic lanterns fed by hexagonal microlenses versus the on sky flat-to-flat field of view per microlens. The spectral resolution is R = 2500, the exposure time is 1800 s and the sky surface brightness is assumed to be 500 photon $s^{-1}$ $m^{-2}$ $arcsecond^{-2}$ $\mu m^{-1}$.

## 6. CONCLUSION

We have now completed the full optical and mechanical designs of PRAXIS and start building it. Extreme care has been taken in increasing the throughput and reducing all type of backgrounds but especially the thermal background. This is necessary to demonstrate that not only the OH suppression system is working properly as in GNOSIS but also that a higher signal-noise ratio can be obtained. Extended simulations of the whole optical train were performed to optimize the system and ensure that the expected performances can be obtained. Future developments include replacing the single mode fibre system by the new multicore system, extending the capability of the instrument in the J band by replacing the grating and making a new IFU and multicore fibre system, and extending the H band to 1.8 µm but this will most probably require a new spectrograph. Ultimately, a system with 2 dichroics and 3 spectrographs would give the whole spectrum from 0.8 µm to 1.8 µm except for the large absorption bands.